\documentstyle{cupconf}
\input{epsf.sty} 

\ifoldfss
\else
  \ifnfssone
    \newmathalphabet{\mathit}
      \addtoversion{normal}{\mathit}{cmr}{m}{it}
      \addtoversion{bold}{\mathit}{cmr}{bx}{it}
    \newmathalphabet{\mathcal}
      \addtoversion{normal}{\mathcal}{cmsy}{m}{n}
    \else
    \ifnfsstwo
    \fi
  \fi
\fi

\newbox\grsign \setbox\grsign=\hbox{$>$} \newdimen\grdimen \grdimen=\ht\grsign
\newbox\simlessbox \newbox\simgreatbox \newbox\simpropbox
\setbox\simgreatbox=\hbox{\raise.5ex\hbox{$>$}\llap
     {\lower.5ex\hbox{$\sim$}}}\ht1=\grdimen\dp1=0pt
\setbox\simlessbox=\hbox{\raise.5ex\hbox{$<$}\llap
     {\lower.5ex\hbox{$\sim$}}}\ht2=\grdimen\dp2=0pt
\setbox\simpropbox=\hbox{\raise.5ex\hbox{$\propto$}\llap
     {\lower.5ex\hbox{$\sim$}}}\ht2=\grdimen\dp2=0pt

\def\a218{\alpha_{2-18}}
\def\aiso{\alpha_{\rm iso}}
\def\aaniso{\alpha_{\rm aniso}}
\def\alphint{\alpha_{\rm int}}
\def\aint218{\alpha_{{\rm int},2-18}}
\def\elldiss{\ell_{\rm diss}}
\def\eps{\epsilon}

\def\Ldiss{L_{\rm diss}}
\def\Lh{L_h}
\def\Ldiss{L_{\rm diss}}
\def\Ls{L_s}
\def\NH{N_{\rm H}}

\def\sigmat{\sigma_{\rm T}}
\def\taut{\tau_{\rm T}}
\def\Te{T_{\rm e}}
\def\Tbb{T_{\rm bb}} 
\def\hexnumber#1{\ifcase#1 0\or1\or2\or3\or4\or5\or6\or7\or8\or9\or
 A\or B\or C\or D\or E\or F\fi }

\makeatletter
\ifx\CUP@mtlplain@loaded\undefined
\else
\fi
\makeatother
 \makeatletter
 \ifx\CUP@mtlplain@loaded\undefined
   \font\tenbmi=cmmib10 at 10pt
   \font\sevenbmi=cmmib10 at 7pt
   \font\fivebmi=cmmib10 at 5pt

   \newfam\bmifam
   \textfont\bmifam=\tenbmi
   \scriptfont\bmifam=\sevenbmi
   \scriptscriptfont\bmifam=\fivebmi
   
 \fi
 \makeatother
\ifnfsstwo

\fi
\ifnfssone

\fi
\ifoldfss

\fi

\mathchardef\varLambda="0103

\makeatletter
\ifx\CUP@mtlplain@loaded\undefined
\else
\fi
\makeatother
\makeatletter
\ifx\CUP@mtlplain@loaded\undefined
  \font\tenbms=cmbsy10
  \font\sevenbms=cmbsy10 at 7pt
  \font\fivebms=cmbsy10 at 5pt
  \newfam\bmsfam
  \textfont\bmsfam=\tenbms
  \scriptfont\bmsfam=\sevenbms
  \scriptscriptfont\bmsfam=\fivebms

  \edef\bsy@{\hexnumber\bmsfam}
  \mathchardef\bnabla="0\bsy@72
\fi
\makeatother
\def\etal{\mbox{\it et al.}}

\title[X-rays and gamma rays from AGN]{X-rays and gamma rays from 
active galactic nuclei}

\author[R. Svensson]%
{R\ls O\ls L\ls A\ls N\ls D\ns  S\ls V\ls E\ls N\ls S\ls S\ls O\ls N}

\affiliation{Stockholm Observatory, S-133 36 Saltsj\"obaden, Stockholm, 
Sweden}

\begin{document}
\ifnfssone
\else
  \ifnfsstwo
  \else
    \ifoldfss
      \let\mathcal\cal
      \let\mathrm\rm
      \let\mathsf\sf
    \fi
  \fi
\fi

\maketitle

\begin{abstract}
Various types of active galactic nuclei (AGN) are briefly discussed, with
an emphasis on the theory of recent X-ray and $\gamma$-ray observations of 
the subclass, Seyfert 1 galaxies. The large radiation power
from AGN is thought to originate from gravitational power released
by matter accreting onto a supermassive black hole.
The physical mechanisms responsible for the energy release and the geometry
of the gaseous components are still uncertain in spite of three 
decades of observational and theoretical studies. 
The recent X-ray and $\gamma$-ray observations, however, start
to provide useful constraints on the models.
This kind of interpretation of the observations is possible
due to theoretical developments during the last decade  of radiative
transfer of X-rays in both optically thick and thin media of various
geometries.
Particular attention is paid to various accretion disk-corona models.
Recent work on exact radiative transfer in such
geometries are reviewed. 
\end{abstract}

\firstsection 

\section{Theoretical history}
\label{sec:theoryhistory}

The underlying framework for almost all efforts to understand
active galactic nuclei (AGN) is the accretion disk picture 
described in the classical papers by Novikov \& Thorne (1973) and \cite{ss73}.
This original picture was partly inspired by the extreme
optical AGN luminosities.
Here, effectively optically thick, rather cold matter forms a 
geometrically thin,
differentially rotating Keplerian disk around a supermassive 
black hole. The differential motion causes viscous
dissipation of gravitational binding energy
resulting in outward transportation of angular momentum and 
inward transportation of matter.
The dissipated energy diffuses vertically and emerges as black
body radiation, mostly in the optical-UV spectral range for the case of AGN. 

Observations have also been the driving force in
the discovery of two other solution branches.

a) Hard X-rays from the galactic black hole candidate, Cyg X-1,
as well as the discovery of strong X-ray emission from most AGN
led to the need for a hot accretion disk solution.
\cite{sle76} (SLE) found  
a hot, effectively optically thin, rather geometrically thin solution 
branch, where the ions and 
the electrons are in energy balance, with the ions being heated 
by dissipation and cooled
through Coulomb exchange, leading to ion temperatures of order 
$10^{11}$ - $10^{12}$ K. 
The efficient cooling of electrons through a variety of mechanisms
above $10^9$ K, leads to electron temperatures being locked
around $10^9$ - $10^{10}$ K. The SLE-solution is thermally unstable.

b) Observations of Cyg X-1 and of radio galaxies
led to two independent discoveries of the third solution branch.
\cite{ich77} developed a model to explain the soft high and 
hard low state of Cyg X-1. The soft high state is due to the disk being
in the optically thick cold state, and the hard low state
occurs when the disk develops into a very hot, optically thin state. 
This solution branch is similar to  
the SLE-solution, except that now the ions are not in local energy
balance. It was found that if the ions were sufficiently
hot they would not cool on an inflow time scale, but rather
the ions would heat up both by adiabatic compression and viscous dissipation
and would carry most of that energy with them into the black hole.
Only a small fraction would be transferred to the electrons,
so the efficiency of the accretion is much less than the
normal $\sim 10 \%$.
As the ion temperature was found to be close to virial, these disks are 
geometrically  thick. Just as for the SLE-case above, 
the electrons decouple to be locked at  $10^9$ - $10^{10}$ K.
The flows resemble the quasi-spherical dissipative flows studied
by, e.g., \cite{mes75} and \cite{mar82}. 
Independently, \cite{rees82} and \cite{phi83} proposed that a similar
inefficient accretion disk solution is responsible for the low 
nuclear luminosities in radio galaxies with large radio lobes and thus 
quite massive black holes. These 
geometrically  thick disks were named {\it ion tori}.  
\cite{rees82} specified more clearly than \cite{ich77}
the critical accretion rate above which
the flow is dense enough for the ions to cool on an inflow time scale
and the ion tori-branch would not exist. Further
considerations of ion tori were made by \cite{beg87}. 
\cite{ich77} emphasized that his version of ion tori is thermally stable.

The ion-tori branch has been extensively studied 
and applied over the last two years (1995-1996) with more than 30 papers
by Narayan and co-workers, Abramowicz and co-workers, as well as many others. 
These studies confirm and extend the original results, 
although some papers do not quote or recognize the original results.
New terminology has been introduced based on an Eularian viewpoint
rather than a Lagrangian. Instead of the ions not cooling,
it is said that the local volume is "advectively cooled" due to the ions
carrying away their energy. The {\it ion tori} are therefore
renamed as {\it advection-dominated disks}. 

The three accretion disk branches were "unified"
by \cite{chen96} and Bj\"ornsson \etal\ \/(1996), who determined
where in parameter space the branches merges.
Parallel to these developments there have been several other lines
of research regarding the physics needed for realistic modelling
of accretion flows. These areas include radiative processes in hot plasmas,
radiation transfer in hot plasmas, radiation transfer of X-rays in cold
plasma, MHD in differentially rotating gas, the origin of disk viscosity,
magnetic flares, gas or MHD-simulations of flows, and so on.
Some of these research lines have already merged. Here we focus
on the areas most important for interpreting the high energy spectra
of AGN. 

While the classical rates for bremsstrahlung, cyclo/synchrotron radiation,
and Compton scattering in the nonrelativistic and relativistic limits
were sufficient in the early disk models, the electron temperatures
of $10^9$ - $10^{10}$ K indicated by both observations and theory
required the calculation of transrelativistic rates of the above processes
as well as for pair processes that becomes important at these
temperatures. Rate calculations as well as exploring the properties
of pair and energy balance in hot plasma clouds
 were done in the 1980s by Lightman, Svensson, Zdziarski and
others. Methods  to incorporate these sometimes complex hot plasma solutions 
into the simplest accretion disk models were developed by
Bj\"ornsson \& Svensson (1991, 1992).
The Compton scattering kernel probably received its definite
treatment in \cite{nag94}.

To obtain approximate spectra, the Kompaneets equation with 
relativistic corrections and with a simple escape probability replacing the
radiative transfer is sufficient (Lightman \& Zdziarski 1987).
However, if one want to obtain constraints on the geometry from
detailed observed spectra, then methods to obtain exact radiative 
transfer/Comptonization solutions in different accretion geometries 
must be developed.
Here again, Igor Novikov had great influence as he and his group stimulated
a particle physicist  and expert on Monte Carlo simulations,
Boris Stern,
to develop a Monte Carlo code able to treat all the high energy 
radiative processes occurring in AGN. The first results
were reported in \cite{kar86} and \cite{nov86}, and an advanced version
of the code was finally documented in \cite{ste95a}.
Faster methods were developed by
\cite{haa93} (approximate treatment of the Compton scattering)
and \cite{pout96} (exact treatment) who solved the radiative transfer
for each scattering order separately (the iterative scattering method).
These codes are fast enough to be implemented in XSPEC, the standard X-ray
spectral fitting package, and exactly computed model spectra can now be used
when interpreting the observations.
So far these methods have been used to study radiative transfer
in two-phase media consisting of cold and hot gas with simple geometries,
but they have yet to be integrated in the  standard accretion disk models. 

UV and X-ray observations indicate the coexistence if both cold and 
hot matter in AGN. Then one must be able to compute the reprocessing
(both absorption, reflection, and transmission) of X-rays by the cold matter. 
The first approximate considerations of Compton reflection by
\cite{gui88} and \cite{whit88} have now led to quite accurate
methods developed by \cite{mag95} and \cite{ponasv96}. Such Compton reflection
forms an important ingredient in the radiative transfer calculations
in two-phase media.

\section{Observational history}
\label{sec:obshistory}

The exploration of the spectra of AGN started some 30 years
ago. It is mainly due to the  developments in instrumental and satellite
technology that we now in the 1990s start having a full broad 
band picture of these spectra. In 1960s groundbased observations
started exploring the optical and radio properties.
The advent of X-ray astronomy allowed the 2-10 keV properties to
be explored in the 1970s. The theoretical modelling then consisted
of connecting the radio, optical, and the X-ray data points with
one or two power laws and discussing the resulting fits within 
nonthermal models where relativistic power law electrons generate
both power law synchrotron radiation and power law  Compton-scattered
radiation. 
The infrared spectral gap was covered in the 1980s.
First after the launch in April 1991 of the {\it Compton Gamma Ray 
Observatory} ({\it CGRO}) 
covering the gamma ray spectral range did we get a full broad band 
view of AGN spectra and thus reliable estimates of their total luminosities
and the spectral ranges where most of the luminosity was emitted.
As described in \cite{der95}, AGN fell into two distinct classes, 
the $\gamma$-loud and the $\gamma$-weak AGN. 
The $\gamma$-loud AGN are the luminous blazars 
where we are looking down into a relativistic jet emerging from the AGN.
Here, the power emerges in two broad band humps, one due to
nonthermal synchrotron emission peaking
in the IR-optical spectral range, the second due to Compton
scattering peaking in the MeV-GeV spectral range.  
The $\gamma$-weak AGN, on the other hand, are the less luminous
Seyfert and radio galaxies (being spiral and ellipticals, respectively), 
where the spectra extends from 
radio frequencies to the soft $\gamma$-rays (100s of keV).
These spectra have approximately equal power per logarithmic 
frequency interval, but show many spectral features, in particular,
three broad band spectral peaks at infrared, UV,
and hard X-ray frequencies.
The fast time variability of the UV to soft $\gamma$-ray spectra
indicates that this emission is "nuclear" radiation originating in the accretion
flow very close to the black hole. The slowly varying broad band 
infrared spectral peak originates at larger distances, most likely being
due to nuclear radiation being reprocessed by dust.
The nuclear spectra has been studied for more than 20 years,
yet little is known about the physical conditions close to the 
super massive black holes in AGN.

Both Seyfert (normally being radio-weak) and radio galaxies comes
in two types based on optical classification. Seyfert 1 and the broad line
radio galaxies show Doppler-broadened  broad (up to 10 000 km/s)
and narrow (up to perhaps 1000 km/s) emission lines,
while the Seyfert 2 galaxies and the narrow line radio galaxies
show only the narrow emission lines. The broad lines originate
in fast-moving photo-ionized $10^4$ K gas clouds within 1 parsec of 
the central source, while the narrow lines originate
in slower moving clouds at kiloparsec distances.
 Aspects of radio galaxies are discussed in the chapter by 
Malcolm Longair in this volume.
Here, we will focus on the Seyfert galaxies only.

In the unified scheme for Seyfert galaxies (e.g. Antonucci 1993), 
we are directly viewing the central X-ray source and broad
line region in Seyfert 1 galaxies, but our line-of-sight
passes through an optically opaque molecular torus in 
Seyfert 2 galaxies obscuring both the central source and the broad line
region but not the narrow line region. The photo-ionized
narrow line gas lies in an ionization cone whose half-opening
angle is determined by the geometry of the obscuring torus.
In some Seyfert 2 galaxies we still see weak polarized broad lines
being reflected by electrons in the ionization cone.
In standard accretion disk scenarios
for the central X-ray source, the unified models imply
that our viewing angle (i.e. the angle between
the disk normal and the line-of-sight) is less than the
half-opening angle of the ionization cone for the case of Seyfert 1s.
Half-opening angles  inferred
from observations are typically 30-40 degrees, which means that
in Seyfert 1s we are viewing the accretion disk from directions
that are closer to face-on than to edge-on.

\section{X-ray and $\gamma$-ray spectra of Seyfert galaxies}
\label{sec:observations}

The satellites (launch year and energy range in parenthesis)
{\it ROSAT} (June 1990; 0.1-2 keV), {\it ASCA} (Feb 1993; 0.1-10 keV),
{\it Ginga} (Feb 1987 - Nov 1991; 2-40 keV) , {\it SAX} (April 1996; 0.2-200 keV), 
{\it XTE} (Dec 1995; 2-200 keV), and {\it CGRO} (April 1991; 50 keV-30 GeV)
have allowed broad band X/$\gamma$-ray studies of some 30 Seyfert galaxies
in the 1990s. However, as AGN varies in the X-rays on time-scales
down to days or even hours simultaneous measurements are necessary. 
Such simultaneous observations unfortunately exist for only a handful
objects, but the number is increasing. 
The launch of {\it INTEGRAL} (2 keV-10 MeV) in the year 2001
will finally provide a large number of truly simultaneous
X/$\gamma$-spectra of AGN.

\begin{figure} 
\leavevmode
\epsfysize=6.3cm  \epsfbox{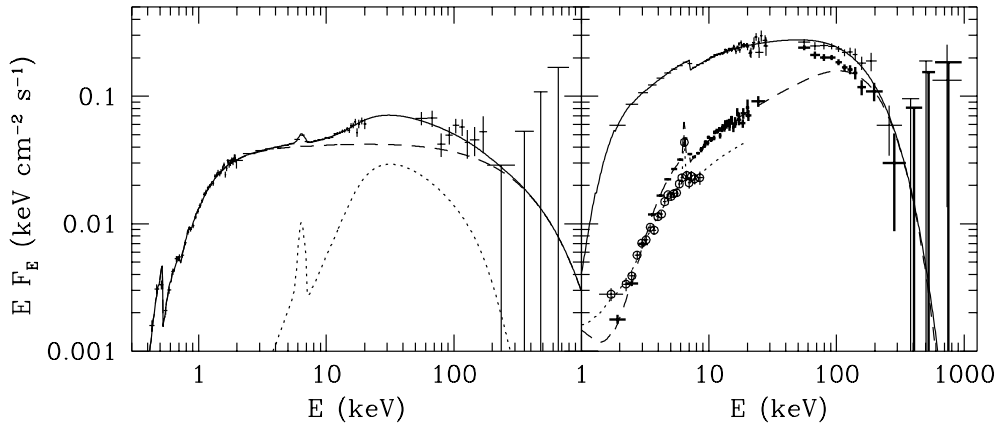}
       \caption{ {\it Left} panel shows
the observed broad band spectrum ({\it crosses}) of IC 4329A from {\it ROSAT},
{\it Ginga}, and {\it CGRO} OSSE (Madejski \etal\ \/1995). The {\it dashed curve} shows the
intrinsic model spectrum incident on both the cold matter and the observer, 
the {\it dotted curve} is the reflected component, 
and the {\it solid curve} is the observed sum. From Magdziarz \&
Zdziarski (1995).
{\it Right} panel shows the observed broad band spectrum  
of NGC 4151  from  {\it Ginga}, and {\it CGRO} OSSE ({\it crosses})
and from {\it EXOSAT} ({\it circled crosses}). The {\it EXOSAT}
data are from April 1984, the highest and lowest {\it Ginga} data from
May 1990 and May 1987, respectively, and the highest and lowest
{\it CGRO} OSSE data from May and April 1993, respectively.
 {\it Solid}, {\it dashed},
and {\it dotted curves} show fits to the  data using a simple thermal
Comptonization model. From 
Zdziarski, Johnson \&  Magdziarz (1996).
              }
         \label{fig1}
   \end{figure}

By combining observations in different energy bands, a picture has emerged 
where the overall shapes of the X-ray and $\gamma$-ray spectra of
different Seyfert 1 galaxies are similar.
The only Seyfert 1 galaxy for which the broad band X- and $\gamma$-ray
spectrum is well determined is IC 4329A (Madejski \etal\ \/1995) being the
second brightest Seyfert galaxy in the 2-10 keV range. 
Simultaneous {\it ROSAT} and {\it CGRO} OSSE, as well as nonsimultaneous
{\it Ginga} data are shown in  Figure~\ref{fig1} ({\it left panel}). 
The observations were modelled by 
Madejski \etal\ \/(1995), Zdziarski \etal\ \/(1994), and  
Magdziarz \& Zdziarski (1995). The spectrum consists of two components:
1) an intrinsic power law  of slope $\alphint 
\approx 0.9$ with an exponential cutoff
energy of about 300 keV (dashed curve in  Fig.~\ref{fig1}), 
and 2) a reflection component caused by 
cold reflecting matter subtending a solid angle $\sim 1-2 \pi$ 
as viewed from the X-ray source (dotted curve in  Fig.~\ref{fig1}).
The reflection component consists of a fluorescent Fe line at 6.4 keV
and a broad peak at about 30 keV 
causing the overall 2-18 keV
spectral slope to be considerably flatter, $\a218 \approx 0.7$ and
the apparent cutoff energy to be smaller, $\approx 50$ keV.
A small column depth of neutral material about 
$\NH = 3 \times 10^{21}$ cm$^{-2}$
 along the line of sight causes absorption below 2 keV.

The average X/$\gamma$-spectrum for four Seyfert 1s using nonsimultaneous
 {\it Ginga} and OSSE data 
shows similar properties as the
spectrum for IC 4329A (Zdziarski \etal\ \/1995).
The same holds for the nonsimultaneous average {\it EXOSAT}/OSSE
spectrum for seven Seyfert 1s 
(Gondek \etal\ \/1996). 

Larger samples exist within narrower spectral ranges using
a single experiment. These samples show spectra consistent
with the smaller samples above.
For example, 60 {\it Ginga} spectra of 27 Seyferts
show $\aint218 \approx 0.95$ and the overall $\a218 \approx 0.73$,
each having a dispersion of about 0.15 (Nandra \& Pounds 1994).

The brightest Seyfert galaxy, the Seyfert 1.5 NGC 4151, once considered 
a prototype for Seyfert 1s, has now turned out to have unusual X-ray
properties. Figure~\ref{fig1} ({\it right panel}) from \cite{zdz96}
shows the extreme spectral states
of NGC 4151. In the 2-18 keV {\it Ginga}-range, the spectral index, 
$\a218$, varies in the range 0.3-0.8 instead of having the
canonical Seyfert value of 0.7. The harder the spectrum is,
the weaker is the 2-18 keV flux, and the spectrum seems to pivot around
100 keV. And indeed, OSSE observations during  
1991-1994 show that the OSSE flux is essentially constant. 
NGC 4151 shows no or a smaller reflection component than
normal Seyfert 1s. This causes the deduced intrinsic power law component
to have a smaller cut off energy than the standard Seyfert 1,
although the total OSSE spectra are indistinguishable.
NGC 4151 also have a larger column depth $\NH \sim 10^{23}$ cm$^{-2}$
of absorbing gas along the line of sight compared to typical Seyfert 1s.
This causes absorption below about 6 keV.
Optical obervations of an ionization cone indicates that we are not viewing
NGC 4151 face-on but rather at a viewing angle of 65 degrees. 
NGC 4151 is clearly a freak object
which requires special consideration to fit into the unified scheme
of Seyfert galaxies. The standard interpretation is that we have a 
direct view of the central X-ray source, but it has also been proposed
(Poutanen \etal\ \/1996) that the central X-ray source is obscured and that the
observed X-ray spectrum is a standard Seyfert 1 spectrum scattered 
by about 65 degrees into the line of sight. Compton recoil 
then decreases the spectral cut off energy.

\begin{figure} 
\leavevmode
\epsfysize=8cm  \epsfbox{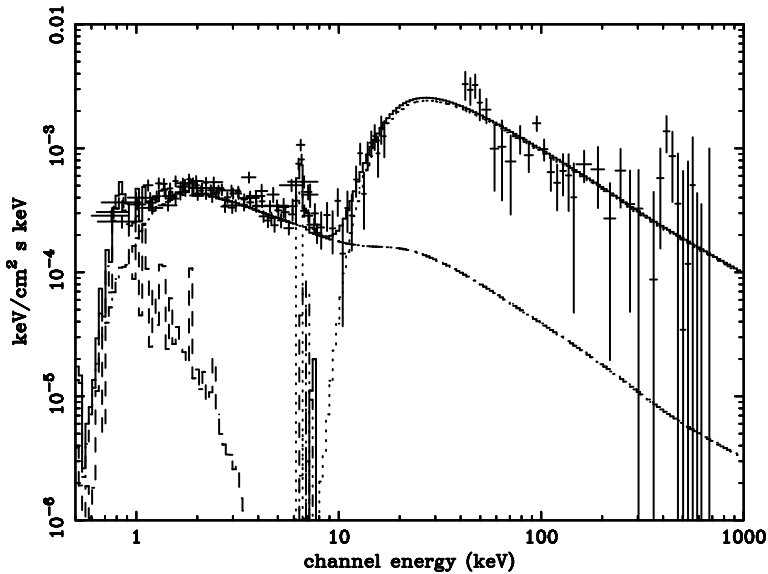}
       \caption{ 
The broad band spectrum ({\it crosses}) of NGC 4945 from {\it ASCA},
{\it Ginga}, and {\it CGRO} OSSE. The {\it dotted curve} is
the direct power law spectrum being absorbed below 20 keV by
a column depth  $\NH \sim 4 \times 10^{24}$ cm$^{-2}$.
The {\it dash-dotted curve} shows the component scattered into the 
line-of-sight
by electrons in the ionization cone. The effects of a 
reflection component is included. From Done, Madejski \& Smith (1996).
               }
         \label{fig2}
   \end{figure}

The observed obscuring column depths in Seyfert 2 galaxies
lies in the range $\NH \sim 2 \times 10^{22}$ -- $5 \times 10^{23}$
cm$^{-2}$ (Smith \& Done 1996). An extreme case is NGC 4945
(Done, Madejski \& Smith 1996) whose
broad band X-ray spectrum is shown in  Figure~\ref{fig2}.
Here, the absorbing column is very large, $\NH \sim 4 \times 10^{24}$ cm$^{-2}$,
causing absorption of the direct spectrum  below 20 keV. 
Below 10 keV the spectrum is dominated by a weak component that has
been scattered into the line-of-sight by electrons  in the ionization
cone. Although weak in the 2-10 keV range, NGC 4945 is the second brightest 
known Seyfert galaxy in the sky above 50 keV.
In the Seyfert 2 galaxy, NGC 1068, the absorbing column is even
larger and the scattered component dominates at least up to 20 keV
(Koyama \etal\ \/1989).

\section{Outstanding questions regarding the X-ray emitting regions
in Seyfert 1 galaxies}
\label{sec:outstanding}

The obvious question is of course which 
radiation process generates the X-ray continuum.
It is commonly believed that Compton scattering by
energetic electrons, either mildly relativistic thermal electrons
or highly relativistic nonthermal electrons, Compton
scatter soft UV-photons into the X-ray range.

Then there is the question of geometry, i.e. the
spatial distribution 
of hot X-ray generating electrons and of cold 
($T < 10^6$ K) reflecting matter. Possible geometries
include a cold slab surrounded either by plane parallel
coronal slabs of hot electrons, or by coronal patches (active regions)
of unknown geometries.

Finally, there is the question whether it is thermal or nonthermal
electrons or both that account for the X-ray emission.
This question might be answered if the spectrum above a few 
hundred keV was known with certainty, but the best available
spectrum of the brightest typical Seyfert 1, IC 4329A, does not have 
sufficient signal to noise at such energies.
The spectrum of IC 4329A can be fit with both thermal
and nonthermal models (Zdziarski \etal\ \/1994).

The nonthermal models studied in the 1980s 
have several attractive features (for review,
see Svensson 1994). In particular, it was predicted already
in 1985 (Svensson 1986) that the pair cascades in the
nonthermal models give rise to an X-ray spectrum 
with $\aint218 \approx 0.9-1.0$ in contradiction to the
overall slope of $\a218 \approx 0.7$ but in agreement
with the intrinsic spectral spectral slope
later found by {\it Ginga} (e.g. Nandra \& Pounds 1994).

Due to the non-detections by COMPTEL and the high energy
cutoffs indicated by OSSE, attention is now 
focused on the thermal models.
 

\section{Geometry of the X-ray source}
\label{sec:geometry}

In principle, the central X-ray source could have spherical symmetry
with the source of UV-photons being small cold clouds
uniformly distributed throughout the X-ray source.
In this case, radiation models invoking spherical symmetry
would be applicable (e.g., 
Ghisellini \& Haardt 1994, Skibo \etal\ \/1995; Pietrini \& Krolik 1995).
In some models, the central X-ray source is quasi-spherical but the
UV-source is anisotropic. Here, the X-ray source can either be a hot
two-temperature SLE-accretion disk or a two-temperature ion
torus. The UV-source may then be the standard cold 
Shakura-Sunyaev disk located beyond some transition radius.
Another proposed alternative is that the disk is hot at larger
radii but that in the innermost region the cooling times become sufficiently
short for the disk to collapse into a cold UV-emitting disk state.
The resulting anisotropy of the UV-photons breaks the spherical symmetry
in both of these scenarios.
Then detailed radiative transfer calculations of
how the UV-photons penetrate the hot disk are required in order 
to determine the emerging X-ray spectrum of Comptonized UV-photons 
at different viewing angles. 
This problem has not yet been solved.

The most commonly used scenario (mainly because of its simplicity)
is the two-phase disk-corona model
(e.g., Haardt \& Maraschi 1991, 1993) in which a hot X-ray emitting corona 
is located above the cold UV-emitting disk of the canonical 
black hole model for AGNs. The power law X-ray spectrum
is generated by thermal Comptonization of the soft UV-radiation.
About half of the X-rays enters and is reprocessed by the cold disk,
emerging mostly as black body disk radiation in the UV.
Haardt \& Maraschi (1991) emphasized the coupling between the 
two phases due to the reprocessing, as the soft disk photons
influence the cooling of the corona. They showed that nearly all
power must be dissipated in the corona in order to have
$\alphint \sim 0.9-1$.
 A consequence of this is that the soft disk
luminosity, $\Ls$, is of the same order as the hard X-ray luminosity,
$\Lh$. The disk-corona scenario is highly anisotropic as the
UV-photons enter the corona from below only.

Observations show that $\Ls$ may be several times larger than 
$\Lh$, in contradiction to the prediction of the uniform two-phase
disk-corona model. This led Haardt, Maraschi, \& Ghisellini (1994)
to propose a patchy disk-corona model, where the corona consists 
of several localized active regions on the surface of the disk.
Internal disk dissipation results in UV-radiation that leaves
the disk without entering the active regions, thus enhancing 
the observed $\Ls$ relative to $\Lh$. The patchy corona model
is also highly anisotropic as the UV-radiation enters the
active regions from below and through the sides. The active regions
could also be elevated above the disk.

The   full radiative
transfer and Comptonization problem in the these geometries
has recently been solved (Haardt \& Maraschi 1993, Stern \etal\ \/1995b,
Poutanen \& Svensson 1996). 
We now turn to discuss the methods and the results.


\section{Radiative transfer/Comptonization} \label{sec:transfer}

Two different methods have been used  to solve the full
radiative transfer/Comp\-toni\-za\-tion problem in mildly relativistic
thermal plasmas accounting for energy and pair balance as well
as reprocessing by the cold disk (including angular anisotropy and
Klein-Nishina effects).

The first method is based on the Non-Linear Monte Carlo (NLMC) method
developed by Stern (1985, 1988) and described in detail in
Stern \etal\ \/(1995a). The Monte Carlo particles (particles and photons)
make up the background so the Monte Carlo particles can interact
with themselves (e.g. photon-photon pair production) causing
nonlinearity. Any geometry can be treated including 2D and 3D
geometries, but calculations so far (Stern et al 1995b)
have been limited to coronal slabs (1D) or active regions (2D) in the
shape of hemispheres or spheres at different elevations
above the cold disc. Another advantage is the possibility to
divide the region into several zones in order to study the
inhomogeneous distributions of, e.g., temperature and pair density.
The drawback is that each run takes a few hours on a Sun IPX.

The second method is based on the iterative scattering method (ISM),
where the radiative transfer is exactly solved for each scattering order
separately (e.g. Sunyaev \& Titarchuk 1985, Haardt 1994).
The code and its testing are briefly described in
Poutanen and Svensson (1996, see also  Poutanen 1994; Poutanen \& Vilhu
1993). It was shown that the results
of the NLMC and the ISM code are in excellent agreement.
The ISM code is a 1D code but it can also treat quasi-2D radiative transfer
in cylinders/pill boxes and hemispheres. The full Compton scattering 
matrix is used allowing to solve polarized radiative transfer
in thermal relativistic plasmas. Fully angular dependent, polarized
Compton reflection is implemented using a Green's matrix 
(Poutanen, Nagendra \& Svensson 1996). The advantage of the ISM code
is that it takes 10 minutes or less on a Sun IPX.

\section{The Comptonization solution} \label{sec:compsolution}

\begin{figure}
\leavevmode
\epsfysize=8cm  \epsfbox{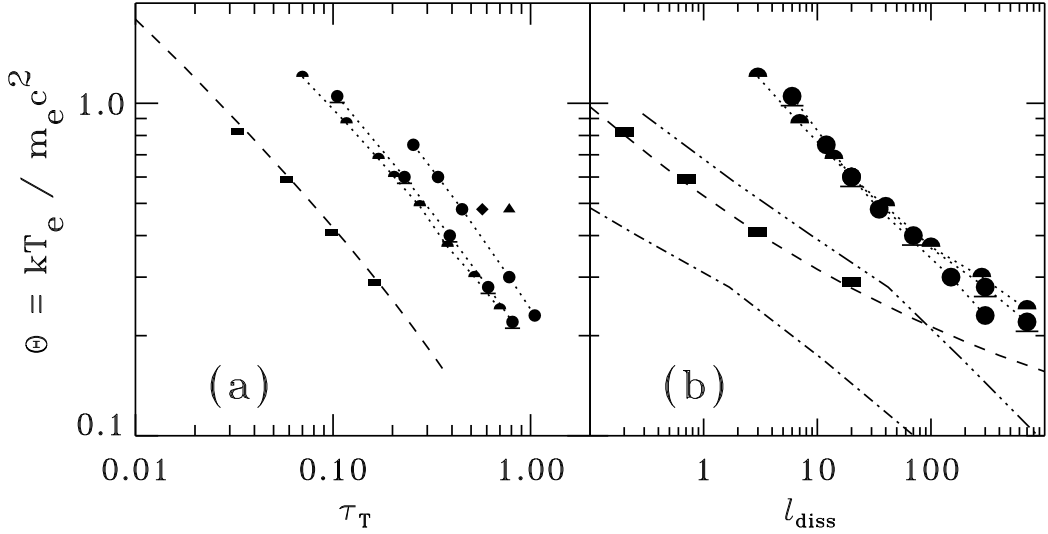}
\caption{Dimensionless volume-averaged temperature, $\Theta \equiv kT_e/ m_e c^2$,
vs. Thomson scattering optical depth, $\taut$, in panel (a), and vs.
 dissipation compactness, $\elldiss \equiv ( \Ldiss /h )$ 
$(\sigmat /  m_e c^3)$ in panel (b), for
a steady X-ray emitting  plasma region in pair and energy
balance on or above a cold disk surface of temperature,  $k\Tbb$ = 5 eV.
The plasma Compton scatters reprocessed soft black body photons from
the cold disk surface. 
{\it Solid rectangles} and {\it dashed curve} show results
from NLMC code and ISM code, respectively, for
the case of a plane-parallel slab corona.  
Results using the NLMC code for individual active pair regions
are shown for
{\it hemispheres} located on the disk surface;
surface spheres also located on the surface ({\it underlined spheres});
spheres located at a height of 0.5$h$ ({\it spheres}),
1$h$ ({\it diamond}), and 2$h$ ({\it triangle}), where $h$ is the radius
of the sphere.
The results for each type of active region are connected by
{\it dotted} curves.  
The {\it dash-dotted} and {\it dash-dot-dot-dotted} curves  in panel (b)
show the critical compactness as function of $\Theta$ above which 
thermalization by Moller and Bhabha scattering is not achieved for the
cases of pair slabs and surface spheres, respectively.
From Stern \etal\ \/(1997).
}
\label{temptaut}
\end{figure}

The two codes have been used to study radiative transfer/Comptonization in pure
pair coronae in energy and pair balance. For coronae of a given geometry
and in energy 
balance, there exists a unique $T_e - \taut$ relation, 
where $\Te$ is the volume-averaged coronal temperature 
and $\taut$ is a characteristic 
Thomson scattering optical depth of the coronal region.
For slabs, $\taut$ is the vertical Thomson scattering optical depth.
For hemispheres and for spheres, the radial $\taut$ is averaged over $2\pi$ solid
angle for hemispheres, and $4\pi$ solid angle for spheres.

In Figure~\ref{temptaut}a, 
this relation is shown for different geometries.
The results for slabs are shown by {\it rectangles}, 
for hemispheres located on the disk surface by {\it hemispheres}, 
for surface spheres located on the disk surface by {\it underlined spheres},
and for spheres located at heights, 0.5$h$, 1$h$, and 2$h$, by {\it spheres},
{\it diamonds}, and {\it triangles}, respectively 
($h$ is the radius of the sphere).
The results for active regions are connected by {\it dotted} curves.
For comparison we also show the slab results from Stern \etal\ \/1995b
 using the ISM code ({\it dashed curve}).
Each curve is characterized by an almost constant generalized
Kompaneets parameter, $y \equiv \taut (1 + \taut) (4 \Theta + 16 \Theta^2)$,
where  $\Theta \equiv kT_e/ m_e c^2$, and where $y \sim 0.49$ for slabs,
$\sim 2.0$ for hemispheres, and $\sim 2.3$, 3.7, 5.9, and 7.8 for the
spheres.
The larger the soft photon starvation (i.e. the fewer the number of 
reprocessed soft photons
reentering the coronal region), the larger is $y$. 
There is very good agreement  
between the slab results from the NLMC and the ISM codes,
which tests the accuracy of both codes and methods.

\section{Pair balance and the compactness}
\label{sec:pairbalance}

Solving the pair balance for the combinations of ($\Theta$, $\taut$)
obtained in \S~\ref{sec:compsolution} gives a unique dissipation
compactness, $\elldiss$ (see Ghisellini \& Haardt 1994 for a discussion).
Here, the local dissipation compactness,
$\elldiss \equiv ( \Ldiss /h ) (\sigmat /  m_e c^3)$,
characterizes the dissipation with $\Ldiss$ being the power providing uniform
heating in a cubic volume of size $h$ in the case of a slab of height $h$, 
or in the whole
volume in the case of an active region of size $h$.
The Figure~\ref{temptaut}b shows the 
$\Theta$ vs. $\elldiss$ relation obtained with the NLMC code
for different geometries and for $k\Tbb$ = 5 eV.
The slab results with the ISM codes are also shown by the dashed curve.  
The parameter space to the right of respective curve is 
forbidden as pair balance cannot be achieved, and  the parameter space
to the left would contain solutions where the background 
coronal plasma dominates over the pairs, i.e. ``pair free'' solutions 
(e.g. Svensson 1984, HM93). Note that the pair temperatures are locked
in the range $\Theta \sim 0.15-1$ (80 -- 500 keV) for the considered
compactnesses. The pairs thus act as a ``thermostat'' (e.g., Svensson 1984)

From Figure~\ref{temptaut}b we find that at a 
given $\Theta$
the active regions have a larger $\elldiss$ than the slabs.
This is due to the longer escape times in slabs.
In a local cubic volume in slabs, the four vertical surfaces act
effectively as ``reflecting'' surfaces increasing the radiation 
energy density in the volume
as compared to the active regions where the radiation escapes freely through
the surface of the volume.

\section{Thermalization}
\label{sec:thermalization}

The question arises whether the electrons can thermalize or not
for the conditions, $\Theta$, $\taut$, and $\elldiss$, in 
Figure~\ref{temptaut}. Energy exchange and thermalization
through Moller ($e^{\pm}e^{\pm}$)
and Bhabha ($e^+e^-$) scattering compete with various loss mechanisms,
with Compton losses being the most important for our conditions.
The thermalization is slowest and the Compton losses largest for the
higher energy particles in the Maxwellian tail.
\cite{ghi93} compared average Maxwellian time scales to find the conditions
when thermalization is not achieved. Instead we use the detailed simulations
by \cite{der89} (their Fig. 8) to find the critical compactness 
above which
the deviation at the Maxwellian mean energy is more than a factor $e \approx 2.7$. 
The {\it dash-dotted} and {\it dash-dot-dot-dotted} curves in
Figure~\ref{temptaut}b 
show this critical compactness for slabs and for surface spheres, respectively.
In agreement with \cite{ghi93}, we find that Moller  
and Bhabha scattering cannot compete with Compton losses in our pair
slabs and active regions. On the other hand,
\cite{ghi88} and \cite{ghi90} found that
cyclo/synchrotron self-absorption acts as a very efficient
thermalizing mechanism as long as the magnetic energy density
dominates the radiation energy density, a condition expected
in the corona and in the magnetic flares on the surface of accretion disks.

\section{Anisotropy effects}
\label{sec:anisotropy}

\begin{figure}
\leavevmode
\epsfysize=8cm  \epsfbox{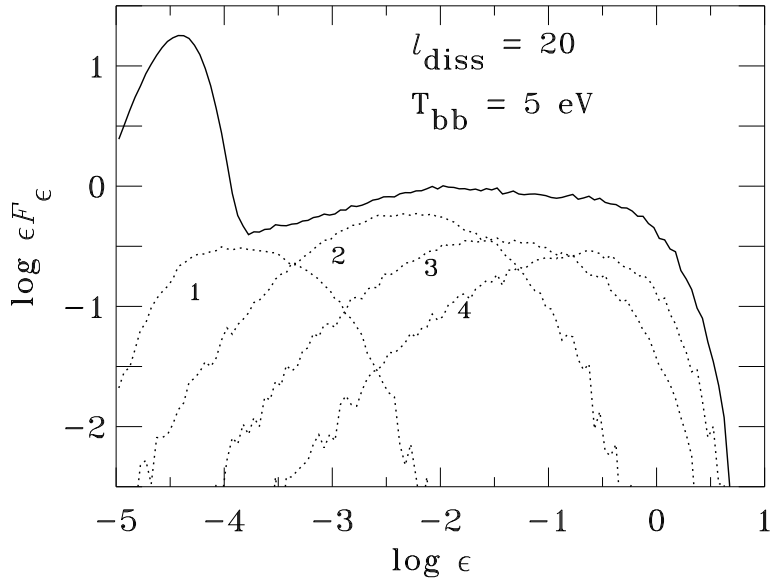}
\caption{Emerging spectrum, $\eps F_{\eps}$, where $F_{\eps}$ is the energy
flux (arbitrary units) and $\eps \equiv h \nu / m_e c^2$ from hemisphere
with compactness, $\elldiss$ = 20, and cold disk temperature, $k \Tbb$ = 5 eV.  
The spectrum is averaged over
viewing angles $0.6 < \cos \theta < 1$.
The {\it  solid curve} shows the sum of the Comptonized spectrum 
from the hemisphere and the reprocessed
black body spectrum (the reflection component is not included).
The numbered {\it dotted curves} show the different scattering orders. 
Note that the first scattering order is significantly reduced due to anisotropic
Compton scattering. This deficiency causes an {\it anisotropy break} close to the
peak of the second scattering order. Below the anisotropy break, the spectrum is
a power law with an {\it anisotropy slope} that is harder than the standard
{\it isotropy slope} above the break.
From Stern \etal\ \/(1997).
}
\label{spectrum}
\end{figure}

Figure~\ref{spectrum} shows the  ``face-on'' emerging spectrum for a hemisphere
with $\elldiss$ = 20 and $k\Tbb$ = 5 eV with the reflected component subtracted.
The {\it solid curve} is the sum of the reprocessed soft black body spectrum and the
Comptonized spectrum. The ``face-on'' spectrum is averaged over 
viewing angles $0.6 < \cos \theta < 1$.
The numbered {\it dotted curves} show spectral profiles of
the different scattering orders. As noted by Haardt (1993) and HM93, the  
first scattering order is significantly reduced in face-on directions due to anisotropic
Compton scattering when $\Theta$ is mildly relativistic. 
This deficiency at low energies causes an {\it anisotropy break} close to the
peak of the second scattering order. Below the anisotropy break, the spectrum is
a power law with an {\it anisotropy slope}, $\aaniso$, that is harder than the standard
{\it isotropy slope}, $\aiso$, above the break. It is the standard isotropy slope, $\aiso$,  
that  has been fitted with several analytical expressions that are functions
of $\Theta$ and $\taut$ (e.g., Zdziarski 1985, Titarchuk 1994). ``Edge-on''
spectra have a much weaker reflection component and no anisotropy break.

\section{Spectra from active pair regions: hemispheres}
\label{sec:hemispectra}

\begin{figure}
\leavevmode
\epsfysize=8cm  \epsfbox{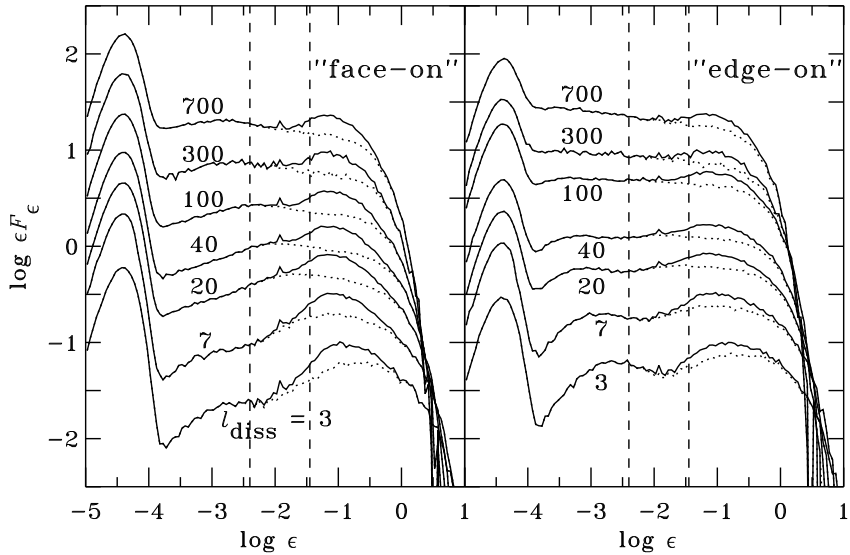}
\caption{Emerging ``face-on'' and ``edge-on'' spectra, $\eps F_{\eps}$, 
where $F_{\eps}$ is the energy
flux (arbitrary units) and $\eps \equiv h \nu / m_e c^2$, from hemispheres
of different compactnesses, $\elldiss$ and for cold disk temperature, $k\Tbb$ = 5 eV. 
The  ``face-on'' spectra are averaged over viewing angles $0.6 < \cos \theta < 1$,
and the  ``edge-on'' spectra over viewing angles $0 < \cos \theta < 0.5$.
The {\it  solid curves} show the total spectrum which is the
sum of the Comptonized spectrum from the hemisphere itself ({\it dotted curves}),
the reprocessed black body spectrum, and the reflection component.
Vertical {\it dashed lines} show the 2-18 keV spectral range. 
The anisotropy break in the face-on spectra moves to lower photon energies 
as $\elldiss$ increases.
From Stern et al. (1995b, 1997).
} 
\label{hemispectra}
\end{figure}

Figure~\ref{hemispectra} shows emerging ``face-one'' and ``edge-on'' spectra
from hemispheres at different $\elldiss$
for $k\Tbb$ = 5 eV.
We have $\aaniso \approx$ 0.77 
while $\aiso \approx$ 1.09. The anisotropy break is therefore about 0.3.
The anisotropy break moves through the 2-18 keV range for $\elldiss \approx$ 10-100,
and thus $\a218$ is smaller than unity for $\elldiss$ less than 100. 
At this low $k\Tbb$ the first Compton order hardly extends into the 2-18 keV range
and there is little  discreteness effect on the behavior of $\a218$.

\section{Diagnostics Using Compactness and {\it Ginga} Slopes}
\label{sec:compginga}

\begin{figure}
\leavevmode
\epsfysize=8cm  \epsfbox{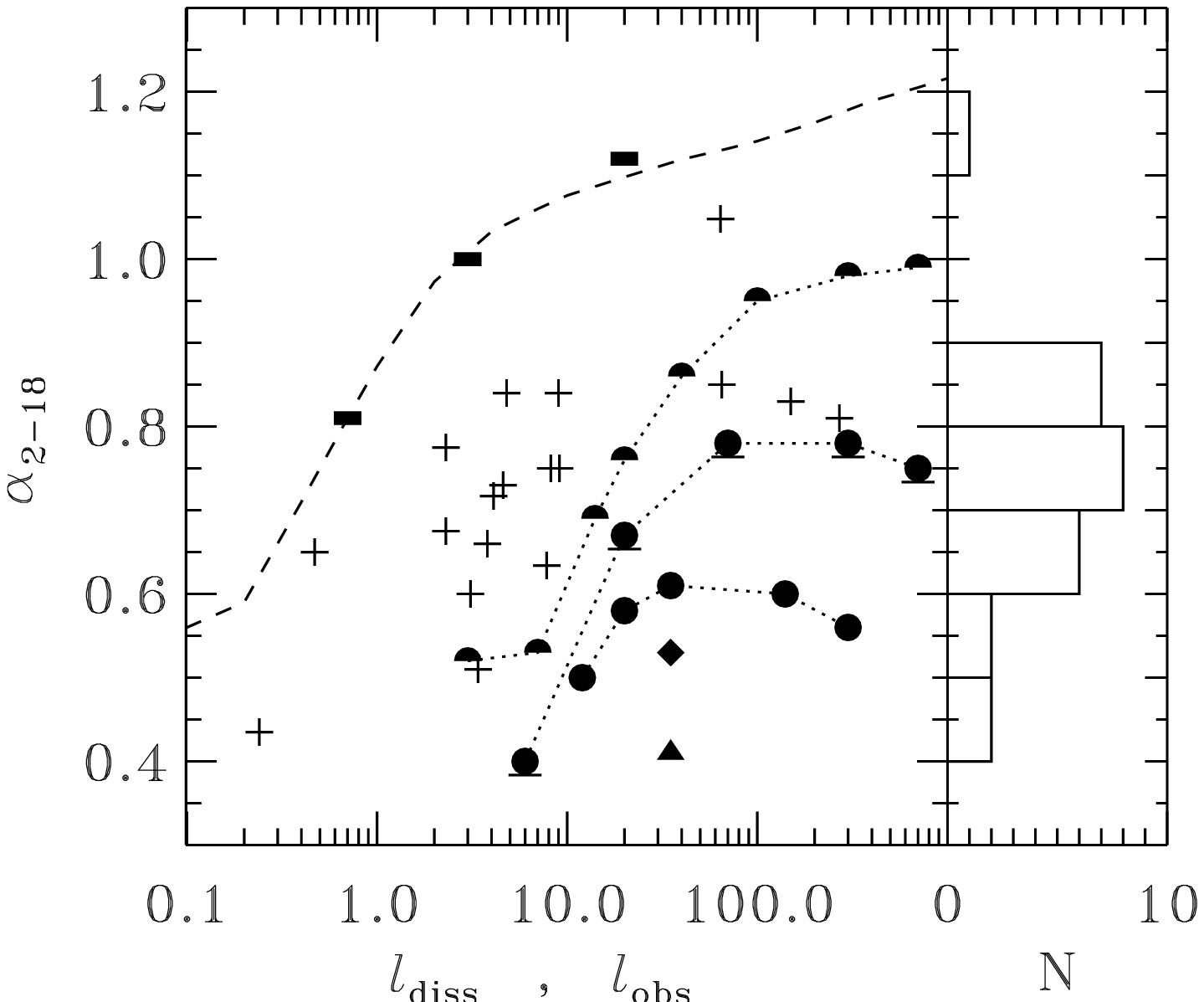}
\caption{Overall spectral intensity index, $\a218$, least square fitted 
to the model spectra in the 2-18 keV range vs. the dissipation compactness,
$\elldiss$. Same notation as in Fig.~3.
The  spectra from the NLMC code were averaged over viewing  angles
$0.6 < \cos \theta < 1.0$ before determining $\a218$ with least square fitting.
For the ISM code, face-on spectra (at $\cos \theta = 0.887$) were used.
The right panel shows the observed distribution of
$\a218$ for {\it Ginga} spectra from 27 Seyfert galaxies (Nandra \& Pounds 1994). 
The crosses represent 17 Seyfert galaxies that have both been observed by 
{\it Ginga}
and have known estimates of their X-ray time variability,
$\Delta t$, and thus of the observed compactness,
$\ell_{\rm obs} \equiv ( L_{\rm obs} /c \Delta t ) 
(\sigmat /  m_e c^3)$.
From Stern \etal\ \/(1995b, 1997).
}
\label{alphaell5ev} 
\end{figure}

The least square overall spectral slope, $\a218$, for the
2-18 keV range were determined and are displayed
in Figure~\ref{alphaell5ev} as a function of the dissipation compactness,
$\elldiss$, for different geometries.
The {\it right panel} of Figure~\ref{alphaell5ev} shows the observed 
distribution of
$\a218$ for {\it Ginga} spectra from 27 Seyfert galaxies (Nandra \& Pounds
1994). The {\it crosses} represent 17 Seyfert galaxies that have both been 
observed by {\it Ginga}
and have known estimates of their X-ray time variability 
(and thus lower limits of their compactnesses). The true crosses may 
lie to the right of the plotted ones. 

One sees that the observations are more consistent with active surface regions
(such as hemispheres or surface spheres) 
than with slabs.  Active regions produce spectra covering
the {\it observed ranges} of $\a218$ ($\approx 0.4-0.9$)  and 
cutoff energies ($\sim 2kT_e$)
for the {\it observed range} of compactnesses.

\section{Conclusions}

Have we been able to throw some light upon the outstanding questions
discussed in \S~\ref{sec:outstanding}?

Regarding the radiative transfer/Comptonization, the progress has been rapid and
it is now possible to use full radiative transfer simulations when fitting
X-ray observations, e.g. with XSPEC, at least for the simplest geometries
(e.g., Poutanen \etal\/ 1996 for the case of NGC 4151).
As anisotropic effects are very important, i.e. as 
spectral shapes depend strongly on viewing angle, it will be possible
to set constraints on the viewing angle. The ISM code also gives
detailed predictions for the spectral X/$\gamma$-ray polarization, 
which may provide crucial constraints once observations become available.
  
The spectra also depend on the geometry of the coronal regions, so
observed spectra can be used as diagnostics of the geometry.
Presently, it seems that active regions are favoured over homogeneous
slab coronae.

There is still no answer to the question whether the radiating electrons
(and positrons) are thermal or nonthermal. High quality spectra above 200 keV
are needed, and the question may not be settled before the 
launch of {\it INTEGRAL} in the year 2001.

\begin{acknowledgments}
      The author acknowledges support from the
      Swedish Natural Science Research Council. 
      He thanks Juri Poutanen, Marek Sikora, and Boris Stern 
      for numerous valuable discussions. Boris Stern's research visits
      at the Stockholm Observatory were financed by the Swedish
      Institute, the Swedish Royal Academy of Sciences, and by
      a NORDITA Nordic project grant.
\end{acknowledgments}


\begin{thebibliography}{}


\bibitem[Antonucci (1993)]{antonucci93}
{\sc Antonucci, R.} 1993 Unified models for active galactic
nuclei and quasars. {\em Ann. Rev. Astr. Astroph.\ \/} {\bf 31}, 
473--521.

\bibitem[Begelman, Sikora \& Rees (1987)]{beg87}
{\sc  Begelman, M. C., Sikora, M. \& Rees, M. J.} 1987
Thermal and dynamical effects of pair production on
two temperature accretion flows. 
{\em Astrophys. J.\ \/} {\bf 313}, 689--698.

\bibitem[Bj\"ornsson \& Svensson (1991)]{bjo91} 
{\sc Bj\"ornsson, G. \& Svensson, R.} 1991 
A recipe for making hot accretion disks.
{\em Astrophys. J. Letters\/} {\bf 371}, L69--L72.

\bibitem[Bj\"ornsson \& Svensson (1992)]{bjo92} 
{\sc Bj\"ornsson, G. \& Svensson, R.} 1992 
Hot pair dominated accretion disks. 
{\em Astrophys. J.\ \/}, {\bf 394}, 500--514.

\bibitem[Bj\"ornsson  \etal\/ (1996)]{bjo96}
{\sc Bj\"ornsson, G., Chen, X., Abramowicz, M. A., \& Lasota, J. P.} 1996
Hot accretion disks revisited.
{\em Astroph. J.\ \/}, in press. 
  
\bibitem[Chen \etal\/ (1996)]{chen96}
{\sc  Chen, X., Abramowicz, M. A., Lasota, J. P., Narayan, R. \& Yi, I.} 1996
Unified description of accretion flow around black holes.
  {\em Astroph. J. Letters } 
{\bf 443}, L61--L64.

\bibitem[Dermer \& Gehrels (1995)]{der95} 
{\sc Dermer, C. D. \& Gehrels, N.} 1995 
Two classes of gamma-ray emitting active galactic nuclei.
{\em Astrophys. J.\ \/} {\bf 447}, 103--120.

\bibitem[Dermer \& Liang (1989)]{der89} 
{\sc Dermer, C. D. \& Liang, E. P.} 1989 
Electron thermalization and heating in relativistic plasmas.
{\em Astrophys. J.\ \/} {\bf 339}, 512--528.

\bibitem[Done, Madejski \& Smith (1996)]{done96} 
{\sc Done, C., Madejski, G. M. \& Smith, D. A.} 1996 
NGC 4945: The brightest Seyfert 2 galaxy at 100 keV.
{\em Astrophys. J. Letters\ \/} {\bf 463}, L63--L66.

\bibitem[Ghisellini \& Haardt (1994)]{ghi94} 
{\sc Ghisellini, G. \& Haardt, F.} 1994
On thermal Comptonization in e$^{\pm}$ pair plasmas.
{\em Astroph. J. Letters\ /} {\bf 429}, L53--L56.

\bibitem[Ghisellini \& Svensson (1990)]{ghi90} 
{\sc Ghisellini, G. \& Svensson, R.} 1990 
Synchrotron self-absorption as a thermalizing mechanism.
in {\em Physical Processes in Hot Cosmic Plasmas}
(eds.\ W.\ Brinkmann, A.\ C.\ Fabian \& F.\ Giovannelli)
pp. 395--400. Kluwer.

\bibitem[Ghisellini, Haardt \& Fabian (1993)]{ghi93} 
{\sc Ghisellini, G., Haardt, F. \& Fabian, A. C.} 1993
On reacceleration, pairs and the high energy spectrum of
AGN and Galactic black hole candidates.
{\em Monthly Not. Roy. Astr. Soc.\ \/} {\bf 263}, L9--L12.
 
\bibitem[Ghisellini, Guilbert \& Svensson (1988)]{ghi88} 
{\sc Ghisellini, G., Guilbert, P. W. \& Svensson, R.} 1988
The synchrotron boiler.
{\em Astrophysical Journal Letters} {\bf 335}, L5--L8.

\bibitem[Gondek \etal\/ (1996)]{gon96} 
{\sc Gondek, D.}, \etal\/ 1996
The average X-ray/gamma-ray spectrum of radio-quiet Seyfert 1s.
{\em Monthly Not. Roy. Astr. Soc.}, in press.
 
\bibitem[Guilbert \& Rees (1988)]{gui88} 
{\sc Guilbert, P. W. \& Rees, M. J.} 1988
`Cold' material in non-thermal sources.
{\em Monthly Not. Roy. Astr. Soc.\ \/} {\bf 233}, 475--484.
 
\bibitem[Haardt (1993)]{haa93} 
 {\sc Haardt, F.} 1993 
Anisotropic Comptonization in thermal plasmas:
spectral distribution in plane parallel geometry. 
{\em Astrophys. J.\ \/} {\bf 413}, 680--693.

\bibitem[Haardt (1994)]{haa94} 
 {\sc Haardt, F.} 1994 
High Energy Processes in Seyfert Galaxies.
PhD dissertation, SISSA, Trieste.

\bibitem[Haardt \& Maraschi (1991)]{hm91} 
{\sc Haardt, F. \& Maraschi, L.} 1991  
A two-phase model for the X-ray emission from Seyfert galaxies.
{\em Astrophys. J. Letters\/} 
{\bf 380}, L51--L54.

\bibitem[Haardt \& Maraschi (1993)]{hm93} 
{\sc Haardt, F. \& Maraschi, L.} 1993 
X-ray spectra from two-phase accretion disks.
{\em Astrophys. J.\ \/} {\bf 413}, 507--517. 

\bibitem[Haardt, Maraschi \& Ghisellini (1994)]{hmg94} 
{\sc Haardt, F., Maraschi, L. \& Ghisellini, G.} 1994
A model for the X-ray and UV emission from Seyfert galaxies
and galactic black holes. 
{\em Astrophys. J. Letters\/} {\bf 432}, L95--L99.

\bibitem[Ichimaru (1977)]{ich77} 
{\sc Ichimaru, S.}  1977 
Bimodal behavior of accretion disks: theory and application to
Cyg X-1 transitions.
{\em Astrophys. J.\ \/} {\bf 214}, 840--855.

\bibitem[Kardashev, Novikov \& Stern (1986)]{kar86} 
{\sc Kardashev, N. S., Novikov, I. D. \& Stern, B. E.} 1986
An electron-positron cauldron and the formation of the hard radiation
of quasars and AGN.  In
{\em IAU Symp. 119: Quasars} 
(ed. G. Swarup \& V. K. Kapahi)
pp. 383--393. D. Reidel. 

\bibitem[Koyama \etal\ \/(1989)]{koy89} 
 {\sc Koyama, K. \etal\ \/} 1989
An intense iron line in NGC 1068.
{\em Publ. Astron. Soc. Japan} {\bf 41}, 731--737.

\bibitem[Lightman \& Zdziarski (1987)]{lz87} 
 {\sc Lightman, A. P. \& Zdziarski, A. A.} 1987
Pair production and Compton scattering in compact sources and comparison
to observations of active galactic nuclei.
{\em Astrophys. J.\ \/} {\bf 319}, 643--661.

\bibitem[Madejski \etal\/ (1995)]{mad95} 
{\sc Madejski, G. M.} \etal\/  1995 
Joint {\it ROSAT-COMPTON GRO} obervations of the X-ray-bright 
Seyfert galaxy IC 4329A.
{\em Astrophys. J.\ \/} {\bf 438}, 672--679.

\bibitem[Magdziarz \& Zdziarski (1995)]{mag95} 
{\sc Magdziarz, P. \& Zdziarski, A. A.} 1995 
Angle-dependent Compton reflection of X-rays and gamma-rays. 
{\em Monthly Not. Roy. Astron. Soc.\ \/} {\bf 273}, 837--848.
 
\bibitem[Maraschi, Roasio \& Treves (1982)]{mar82} 
{\sc Maraschi, L., Roasio, R. \& Treves, A.} 1982 
The effect of multiple Compton scattering on the temperature
and emission spectra of accreting black holes.
{\em Astrophys. J.\ \/} {\bf 253}, 312--317.

\bibitem[M\'esz\'aroz (1975)]{mes75} 
{\sc M\'esz\'aroz, P.} 1975 
Radiation from spherical accretion onto black holes.
{\em Astr. Astroph.\ \/} {\bf 44}, 59--68.

\bibitem[Nagirner \& Poutanen (1993)]{nag93} 
{\sc Nagirner, D. J. \& Poutanen, J.} 1993 
Compton scattering of polarized light: scattering matrix for 
isotropic electron gas.
{\em Astr. Astroph.\ \/} {\bf 275}, 325--336.

\bibitem[Nagirner \& Poutanen (1994)]{nag94} 
{\sc Nagirner, D. J. \& Poutanen, J.} 1994 
Single Compton scattering.
{\em Astrophys. Space Phys.\ \/} {\bf 9}, 1--47. 

\bibitem[Nandra \& Pounds (1994)]{nan94} 
{\sc Nandra, K. \& Pounds, K. A.} 1994 
{\it Ginga} observations of the X-ray spectra of Seyfert galaxies.
{\em Monthly Not. Roy. Astron. Soc.\ \/} {\bf 268}, 405--429. 

  
\bibitem[Novikov \& Stern (1986)]{nov86} 
{\sc Novikov, I. D. \& Stern, B. E.} 1986
A possible mechanism of the formation of the hard spectrum 
of active galactic nuclei.  In
{\em Structure and Evolution of Active Galactic Nuclei} 
(ed. G. Gluricin \etal\/)
pp. 149--171. D. Reidel. 

\bibitem[Novikov \& Thorne (1973)]{nov73} 
{\sc Novikov, I. D. \& Thorne, K. S.} 1973
Astrophysics of black holes. In
{\em Black Holes, Les Houches} (ed. C. De Witt \& B. DeWitt)
pp. 343--450. Gordon \& Breach.

\bibitem[Phinney (1983)]{phi83} 
{\sc Phinney, E. S.} 1983
A theory of radio sources. 
PhD dissertation, University of Cambridge.

\bibitem[Pietrini \& Krolik (1995)]{pie95} 
{\sc Pietrini, P. \& Krolik, J. H.} 1995  
The inverse Compton thermostat in hot plasmas near 
accreting black holes.
{\em Astrophys. J.\ \/} {\bf 447}, 526--544.

\bibitem[Poutanen (1994)]{pout94} 
{\sc Poutanen, J.} 1994 
Compton scattering matrix for relativistic Maxwellian electron
distribution. 
{\em J. Quant. Spectrosc. Rad. Transfer \/}{\bf 51}, 813--822.
 
\bibitem[Poutanen (1994)]{pou94} 
{\sc Poutanen, J.} 1994 
Compton scattering of polarized light in
active galactic nuclei and X-ray binaries.
PhD dissertation, University of Helsinki.

\bibitem[Poutanen \& Svensson (1996)]{pout96} 
{\sc Poutanen, J. \& Svensson, R.} 1996 
The two-phase pair corona model for active galactic nuclei and
X-ray binaries: How to obtain exact solutions.
{\em Astrophys. J.\ \/} in press. 

\bibitem[Poutanen \& Vilhu (1993)]{pout93} 
{\sc Poutanen, J. \& Vilhu, O.} 1993 
Compton scattering of polarized light in two-phase accretion discs.
{\em Astr. Astroph.\ \/} {\bf 275}, 337--344.

\bibitem[Poutanen, Nagendra \& Svensson (1996)]{ponasv96} 
{\sc Poutanen, J., Nagendra, K. N. \& Svensson, R.} 1996
Green's matrix for Compton reflection of polarized radiation 
from cold matter. 
{\em Monthly Not. Roy. Astron. Soc.\ \/}, in press. 

\bibitem[Poutanen \etal\/ (1996)]{poutetal96} 
{\sc Poutanen, J., Sikora, M., Begelman, M. C. \& Magdziarz, P.} 1996
The Compton mirror in NGC 4151.
{\em Astrophys. J. Letters} {\bf 465}, L107--L110.

\bibitem[Rees \etal\/ (1982)]{rees82}
{\sc Rees, M. J., Begelman, M. C., Blandford, R. D. \& Phinney, E. S.}
1982 
Ion-supported tori and the origin of radio jets.
{\em Nature} {\bf 295}, 17--21.

\bibitem[Shakura \& Sunyaev (1973)]{ss73} 
{\sc Shakura, N. I. \& Sunyaev, R. A.} 1973 
Black holes in binary systems. Observational appearance.
{\em Astr. Astroph.\ \/} {\bf 24}, 337--355.

\bibitem[Shapiro, Lightman \& Eardley (1976)]{sle76}
{\sc Shapiro, S. L., Lightman, A. P. \& Eardley D. N.} 1976 
A two-temperature accretion disk model for Cygnus X-1:
structure and spectrum.
{\em Astrophys. J.\ \/} {\bf 204}, 187--199.

\bibitem[Skibo, Dermer, Ramaty \& McKinley (1995)]{ski95} 
{\sc Skibo, J. G., Dermer, C. D., Ramaty, R. \& McKinley, J. M.} 1995
Thermal Comptonization in mildly relativistic pair plasmas.
{\em Astrophys. J.\ \/} {\bf 446}, 86--100.

\bibitem[Smith \& Done (1996)]{smi96} 
{\sc  Smith, D. A. \& Done, C.} 1996 
Unified theories of AGN: A hard X-ray sample of Seyfert 2 galaxies.
{\em Monthly Not. Roy. Astron. Soc.\ \/}, in press.
  
\bibitem[Stern (1985)]{ste85} 
{\sc Stern, B. E.} 1985 
On the possibility of efficient production of electron-positron pairs
near pulsars and accreting black holes.
{\em Sov.Astr.\ \/} {\bf 29}, 306--313.


\bibitem[Stern (1988)]{ste88} 
{\sc Stern, B. E.} 1988
Nonthermal pair production in active galactic nuclei: A detailed radiation
transfer model. 
{\em Nordita/88-51 A}, preprint.

\bibitem[Stern \etal\/ (1995a)]{ste95a} 
{\sc Stern, B. E., Begelman, M. C., Sikora, M. \& Svensson, R.} 1995a
A large-particle Monte Carlo code for simulating non-linear
high-energy processes near compact objects.
{\em Monthly Not. Roy. Astron. Soc.\ \/} {\bf 272}, 291--307.

\bibitem[Stern \etal\/ (1995b)]{ste95b} 
{\sc Stern, B. E., Poutanen, J., Svensson, R., Sikora, M. \& 
Begelman M. C.} 1995b 
On the geometry of the X-ray emitting region in Seyfert galaxies.
{\em Astrophys. J. Letters\/} {\bf 449}, L13--L17. 

\bibitem[Stern \etal\/ (1997)]{ste97} 
{\sc Stern, B. E.} \etal\/ 1997, in preparation.
 
\bibitem[Sunyaev \& Titarchuk (1985)]{sun85} 
{\sc Sunyaev, R. A. \& Titarchuk, L. G.} 1985 
Comptonization of low-frequency radiation in accretion disks:
angular distribution and polarized hard radiation.
{\em Astr. Astroph.\ \/} {\bf 143}, 374--388. 

\bibitem[Svensson (1984)]{sve84} 
{\sc Svensson, R.} 1984 
Steady mildly relativistic thermal plasmas: processes and properties.
{\em Monthly Not. Roy. Astron. Soc.\ \/} 
{\bf 209}, 175--208.

\bibitem[Svensson (1986)]{sve86} 
{\sc Svensson, R.} 1986 
Physical processes in active galactic nuclei.
In  {\em IAU Coll. 89: Radiation Hydrodynamics 
in Stars and Compact Objects} (ed. D. Mihalas \& K-H. Winkler),
pp. 325--345. Springer.

\bibitem[Svensson (1994)]{sve94} 
{\sc Svensson, R.} 1994 
The nonthermal pair model for the X-ray and gamma-ray spectra from
active galactic nuclei.
{\em Astrophys. J. Suppl.\ \/} {\bf 92}, 585--592.

\bibitem[Titarchuk (1994)]{tit94}
{\sc Titarchuk L.} 1994 
Generalized Comptonization models and applications to recent 
high-energy observations.
{\em Astrophys. J.\ \/} {\bf 434}, 570--586.

\bibitem[White, Lightman, \& Zdziarski (1988)]{whit88} 
{\sc White, T. R., Lightman, A. P., Zdziarski, A. A.} 1988 
Compton reflection of gamma rays by cold electrons.
{\em Astrophys. J.\ \/} {\bf 331}, 939--948.


\bibitem[Zdziarski (1985)]{zdz85} 
{\sc Zdziarski, A. A.} 1985 
Power-law X-ray and gamma-ray emission from relativistic thermal
plasmas. 
{\em Astrophys. J.\ \/} {\bf 289}, 514--525.

\bibitem[Zdziarski \etal\/ (1994)]{zdz94} 
{\sc Zdziarski, A. A.} \etal\/ 1994 
Physical processes in the X-ray/gamma-ray source of IC 4329A.
{\em Monthly Not. Roy. Astron. Soc.\ \/} 
{\bf 269}, L55-L60.

\bibitem[Zdziarski, Johnson \& Magdziarz (1996)]{zdz96} 
{\sc Zdziarski, A. A., Johnson, W. N. \& Magdziarz, P.} 1996 
Broad-band gamma-ray and X-ray spectra of NGC 4151 and their
implications for physical processes and geometry.
{\em Monthly Not. Roy. Astron. Soc.\ \/}, in press.
            

\bibitem[Zdziarski \etal\/ (1995)]{zdz95} 
{\sc Zdziarski, A. A., Johnson, W. N., Done, C., Smith, D. \&
McNaron-Brown, K.} 1995 
The average X-ray/gamma-ray spectra of Seyfert galaxies from
{\it Ginga} and OSSE and the origin of the comic X-ray background.
{\em Astrophys. J. Letters\/} {\bf 438}, L63--L66.

 

\end{thebibliography}
\end{document}